\documentclass[pra,aps,amsmath,amssymb,amsfonts,twocolumn,nofootinbib,longbibliography,superscriptaddress]{revtex4-2}
\usepackage{amssymb}
\usepackage{bm,mathrsfs}
\usepackage{graphicx}
\usepackage{epsfig}
\usepackage{amsmath,bbm}
\usepackage{amsfonts,amssymb}
\usepackage{times}
\usepackage{verbatim}
\usepackage[sort&compress]{natbib}
\usepackage{amsmath}
\usepackage{bm}
\usepackage{float}

\allowdisplaybreaks[4]
\usepackage[colorlinks,breaklinks,linkcolor=blue,anchorcolor=blue,citecolor=blue,urlcolor=blue]{hyperref}
\begin{document}

\title{Generation of complete graph states in a spin-$1/2$ Heisenberg chain with a globally optimized magnetic field}

\author{X. X. Li}
\affiliation{Sino-European Institute of Aviation Engineering, Civil Aviation University of China, Tianjin 300300, China}

\author{D. X. Li}
\affiliation{College of Physics Science and Technology, Shenyang Normal University, Shenyang 110034, China}

\author{X. Q. Shao}
\email{shaoxq644@nenu.edu.cn}
\affiliation{Center for Quantum Sciences and School of Physics, Northeast Normal University, Changchun 130024, China}
\affiliation{Center for Advanced Optoelectronic Functional Materials Research, and Key Laboratory for UV Light-Emitting Materials and Technology of Ministry of Education, Northeast Normal University, Changchun 130024, China}

\begin{abstract}
Graph states possess significant practical value in measurement-based quantum computation, with complete graph states that exhibit exceptional performance in quantum metrology. In this work, we introduce a method for generating multiparticle complete graph states using a spin-$1/2$ Heisenberg $XX$ chain subjected to a time-varying magnetic field, which applies to a wide range of systems. Our scheme relies exclusively on nearest-neighbor interactions between atoms, with real-time magnetic field formation facilitated by quantum optimal control theory.
We focus specifically on neutral-atom systems, finding that multiparticle complete graph states with $N=3\sim6$ can be achieved in less than $0.25~\mu{\rm s}$, using a hopping amplitude of ${J}/{(2\pi)} = -2.443~{\rm MHz}$. This assumes an initial state provided by an equal-weight superposition of all spin states that are encoded by the dipolar-interacting Rydberg states. Furthermore, we thoroughly address various experimental imperfections and showcase the robustness of our approach against atomic vibrations, fluctuations in pulse amplitude, and spontaneous emission of Rydberg states. Taking into account the common occurrence of disturbances in the experimental setups of neutral-atom systems, our one-step strategy to achieve such graph states emerges as a viable alternative to techniques based on controlled-Z gates.
\end{abstract}

\maketitle

\section{Introduction}
Graph states, derived from mathematical graph theory \cite{PhysRevA.69.062311,PhysRevA.89.052335}, play a central role in quantum information theory. It describes a class of multiparticle-entangled states including Greenberger-Horne-Zeilinger (GHZ) states and cluster states \cite{PhysRevLett.86.910}. Such states form a universal resource for measurement-based quantum computing (MBQC) without controlled two-system quantum gates
\cite{PhysRevLett.86.5188,PhysRevA.82.030303,nphoton.2010.283,PhysRevLett.112.120504,Reimer2019}. Some of these can also be widely applied in quantum teleportation \cite{PhysRevA.78.042302,ACCESS8653915,Kazemikhah2022} and quantum secret sharing \cite{PhysRevA.78.042309,PhysRevA.82.062315}.
Multiparticle complete graph states represent that any two particles interact with each other, requiring more projective measurements to be disentangled. When a single $\sigma_{y}$ measurement is performed at an arbitrary vertex (qubit), these states can be converted to star graph states (GHZ type) \cite{PhysRevA.69.062311,PhysRevA.69.022316}. Such states exhibit excellent performance in quantum metrology \cite{Toth_2014,PhysRevLett.124.110502}, quantum error-correction encoding \cite{PhysRevA.105.042418}, and quantum cryptography \cite{PhysRevA.105.052420}.
However, although complete graph states have many advantages, less attention has been paid to preparing such graph states experimentally, focusing mainly on linear and 2D cluster states \cite{science.aah4758,Reimer2019,science.aay2645,Yang2022}.

The preparation of multiparticle complete graph states can be conceptually divided into two primary categories. The first approach is applicable to arbitrary graph states, employing the mathematically proven two-qubit controlled-Z gate (CZ) \cite{PhysRevLett.95.160501,PhysRevA.77.032308,Ananth2016}. However, these protocols are constrained by the operational time and fidelity of the logic gate.
On the contrary, complete multiparticle graph states can also be generated within the framework of cavity quantum electrodynamics \cite{PhysRevLett.95.110503,PhysRevA.75.034308,PhysRevA.79.062319}. In particular, certain theoretical schemes \cite{PhysRevA.77.014304,PhysRevA.77.014304,SHAO3132} enable one-step realization, but these one-step approaches require uniform interaction strengths between all particle pairs, which poses challenges for experimental implementation. Recently, the proposal of an equivalent Hamiltonian (Heisenberg $XXX$ model with a staggered field) \cite{PhysRevLett.126.160402} has sparked interest in Floquet engineering for the construction of such a Hamiltonian \cite{science.abd9547,PRXQuantum.3.020303}. Nevertheless, in practice, controlling the impact of noncommutation of the Hamiltonian under periodic driving becomes challenging as the particle number increases.

As quantum control theory progresses, optimal quantum control provides a toolkit to craft external field shapes to perform assigned tasks, such as state generation \cite{PhysRevA.79.042304,Glaser2015,PhysRevA.93.010304,science.aax9743}, quantum gate implementation \cite{PhysRevA.90.032329,PhysRevApplied.17.014036}, and noisy intermediate-scale quantum technology \cite{Preskill2018quantumcomputingin}. These tools aim to achieve the specified objectives with minimal energy and resource expenditure. The practical realization of calculated field shapes has been facilitated by the availability of commercially accessible arbitrary waveform generators.
In this paper, using the Gradient Ascent Pulse Engineering (GRAPE) optimization algorithm \cite{KHANEJA2005296,PhysRevA.99.042327}, we propose an alternative Hamiltonian represented as a spin-$1/2$ Heisenberg $XX$ chain with identical nearest-neighbor interactions under an optimized time-dependent magnetic field. Governed by this Hamiltonian, we successfully generate complete graph states for a particle number ranging from $N=3$ to $N=6$. Our protocol boasts two key advantages: (i) It exclusively relies on the nearest-neighbor interactions, ensuring robust operability and scalability in experimental setups. (ii) The operational time is significantly reduced to less than $0.25~\mu{\rm s}$, a crucial factor in mitigating the effects of decoherence.

The structure of this paper is organized as follows. In Sec.~\ref{II}, we initiate our exploration with the introduction of the definition of the complete graph state and an analysis of the corresponding scheme. Subsequently, in Sec.~\ref{III}, we employ the GRAPE optimization algorithm to articulate a system-independent realization of complete graph states in a ``one-step" fashion. Transitioning to Sec.~\ref{IV}, we delve into a specific physical system, the Rydberg atom array, providing a comprehensive examination of the experiment's feasibility. Our results suggest that, taking into account experimental noise, the population of the complete graph state with $N=6$ can exceed $0.91$ in a time frame of $0.25~\mu{\rm s}$. Finally, Sec.~\ref{V} encapsulates our findings, offering a concise summary.

\section{one-step Generation of complete graph states}\label{II}
\begin{figure}
\centering\scalebox{0.27}{\includegraphics{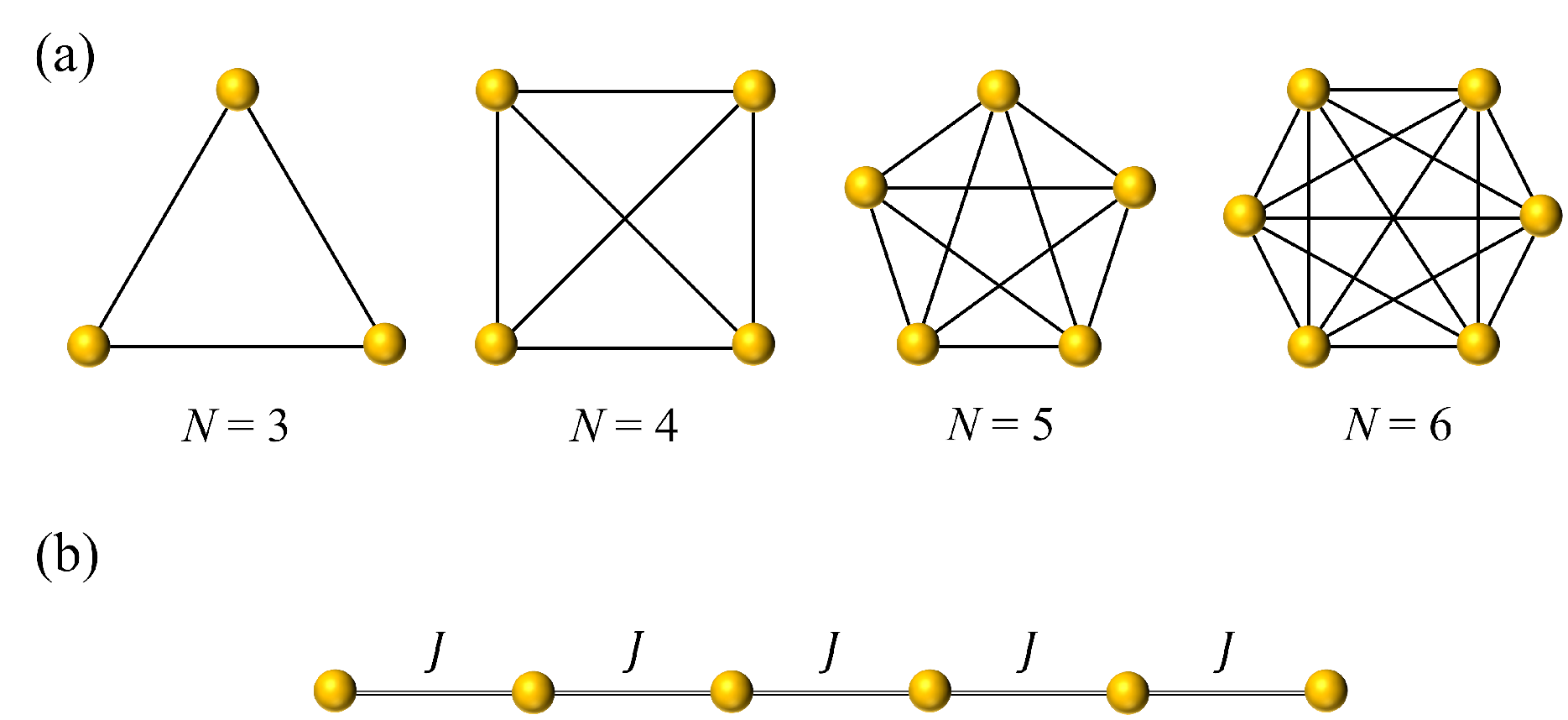}}
\caption{\label{Complete}(a) Multiparticle complete graph states with the number of vertices $N=3\sim6$. The graph connecting the two associated vertices has an edge whenever two particles have interacted. (b) Experimental geometry. Particles are arranged as a chain with the open boundary condition and have equivalent interactions between the nearest particles.}
\end{figure}

Mathematically, a graph can be divided into vertices and edges, $G=\{V,E\}$. Physically, such graphs can constitute the graph states, with the vertices playing the role of physical systems, while the edges represent interactions. As shown in Fig.~\ref{Complete}(a), the multiparticle complete graph states represent that there exists an interaction between two arbitrary qubits. Taking the vertices as spin systems with $\{|\uparrow\rangle,|\downarrow\rangle\}$, the complete graph states with $N$ vertices can be defined as
\begin{equation}\label{target}
|K_{N}\rangle=\frac{1}{2^{N/2}}\bigotimes_{i=1}^{N}[|\uparrow_{i}\rangle(-1)^{N-i}\prod_{j=i+1}^{N}\sigma_{z}^{j}+|\downarrow_{i}\rangle],
\end{equation}
with the convention $\sigma_{z}^{N+1}\equiv1$. To generate such complete graph states, we construct a Hamiltonian that holds a global time-dependent magnetic field, which takes the form of ($\hbar=1$)
\begin{equation}\label{Eq_time}
H(t)=H_{\textrm{XX}}+\sum_{i=1}^{N}B(t)S^{z}_{i},
\end{equation}
where $S^{z}$ is the spin operator, and
\begin{equation}\label{xx}
H_{\textrm{XX}}=\sum_{\langle i,j\rangle}\frac{J}{2}(\sigma^{x}_{i}
\sigma^{x}_{j}+\sigma^{y}_{i}\sigma^{y}_{j})
\end{equation}
is a spin-$1/2$ Heisenberg $XX$ chain model with Pauli operators $\sigma^{x(y,z)}$, where $\langle i,j\rangle$ indicates that the sum is taken over all nearest neighbor pairs $(i, j)$ in the spin chain. Taking $N=3$ as an example, the target state can be written as
\begin{eqnarray}
|K_{3}\rangle&=&\frac{1}{2\sqrt{2}}(|\uparrow\uparrow\uparrow\rangle+|\uparrow\uparrow\downarrow\rangle+|\uparrow\downarrow\uparrow\rangle
-|\uparrow\downarrow\downarrow\rangle\nonumber\\&&+|\downarrow\uparrow\uparrow\rangle-|\downarrow\uparrow\downarrow\rangle
-|\downarrow\downarrow\uparrow\rangle-|\downarrow\downarrow\downarrow\rangle),
\end{eqnarray}
which presents as a superposition of eight states that can be separated into two groups with opposite signs, i.e. $\{|\uparrow\uparrow\uparrow\rangle,|\uparrow\uparrow\downarrow\rangle,|\uparrow\downarrow\uparrow\rangle,|\downarrow\uparrow\uparrow\rangle\}$ and $\{|\uparrow\downarrow\downarrow\rangle,|\downarrow\uparrow\downarrow\rangle,|\downarrow\downarrow\uparrow\rangle,|\downarrow\downarrow\downarrow\rangle\}$.
$H_{\rm xx}$ described by Eq.~(\ref{xx}) has the capability to alter the phase of the state within the associated excitation subspace through flip-flop interactions while ensuring the conservation of excitation numbers. Furthermore, the introduction of a global magnetic field $\sum_{i=1}^{3}B(t)S^{z}_{i}$ allows the accumulation of distinct phases in various excitation subspaces. The synergistic integration of these two factors contributes to the realization of a complete graph state of three particles.

Since the two parts ($H_{\textrm{XX}}$ and the global magnetic field term) commute with each other, the evolution can take place under the Hamiltonian (\ref{Eq_time}) or under the $H_{\textrm{XX}
}$ and the global magnetic field separately. Theoretically, any equivalent $H_{\textrm{XX}}$ model can be applied to this scenario. Under a global magnetic field with constant amplitude, an exact analytical solution of $B$ can be found for $N=3$, as shown in Appendix~\ref{app}. However, for $N>3$, such an exact solution is non-existent. Thus, we consider a time-dependent magnetic field $B(t)$ and introduce the GRAPE optimization algorithm to find the optimal solution for the system. In this way, we can expect to obtain a multiparticle universal protocol.
\begin{figure*}
\centering\scalebox{0.3}{\includegraphics{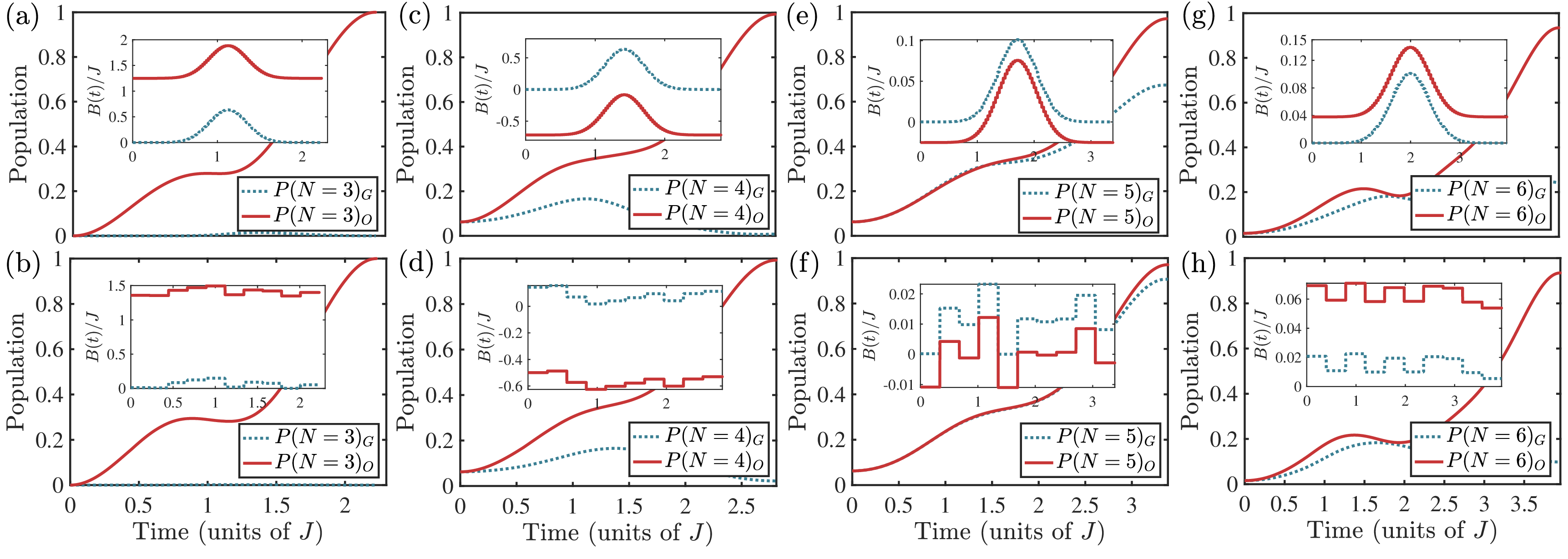}}
\caption{\label{Multi_Per}The realization of the multiparticle complete graph states with $N=3\sim6$ governed by Eq.~(\ref{Eq_time}) initialized from state $|\Psi_{0}\rangle=\bigotimes_{i=1}^{N}[{1}/{\sqrt{2}}(|\uparrow\rangle+|\downarrow\rangle)]$. (a) and (b) respectively show evolution under the magnetic field in random and Gaussian types for $N=3$. The dotted lines correspond to the evolution under the guess field, while the solid lines correspond to the optimized one. Then (c) and (d) show the evolution for $N=4$, while (e) and (f) for $N=5$, and (g) and (h) for $N=6$. The insets show the corresponding time-dependent $B(t)$ of the initial guess and the optimized result.}
\end{figure*}

\section{The utilization of the GRAPE optimization algorithm}\label{III}
Given the known form of the target state $|K_{N}\rangle$, the optimization problem, which is implemented using the gradient descent algorithm, is translated into the search for the optimal amplitudes $B(t)$ of the magnetic field. These amplitudes aim to guide the initial state $\rho(0)$ to $\rho(T)$ with maximum similarity to $\rho_{K}=|K_{N}\rangle\langle K_{N}|$ within a specified time duration $T$. In our protocol, the system Hamiltonian is segmented into two distinct parts.
\begin{equation}
H(t)=H_{\textrm{XX}}+B(t)H_{z},
\end{equation}
where $H_{z}=\sum_{i=1}^{N}S^{z}_{i}$ is referred to as the control component. The quantum control landscape, depicting the expectation value of a Hermitian observable operator $\rho_{K}$ at a given time $T$, is represented as \cite{HO2006226}
\begin{equation}
\Phi[B(\cdot)]=\textrm{Tr}[\rho(0)\rho_{K}(T)],
\end{equation}
where $\rho_K(T)=U^{\dag}(T,0)\rho_{K}U(T,0)$. The variation of the observable operator $\rho_{K}$ is expressed as
\begin{equation}
\delta\rho_{K}(t)=\int_{0}^{t}[\rho_{K}(t'),H_{z}(t')]\delta B(t')dt',
\end{equation}
where $H_{z}(t')=-iU^{\dag}(t',0)H_{z}U(t',0)$. This results in the fundamental equation
\begin{equation}
\frac{\delta\rho_{K}(T)}{\delta B(t')}=[\rho_{K}(T),H_{z}(t')],~~\forall t'<T.
\end{equation}
Consequently, the corresponding gradient is given by
\begin{equation}
g[B(t)]=\frac{\partial \Phi[B(\cdot)]}{\partial B(t')}=\textrm{Tr}[[\rho(0),\rho_{K}(T)]H_{z}(t')].
\end{equation}
Therefore, the performance function $\Phi[B(\cdot)]$ can be enhanced by selecting
\begin{equation}
B(k)\rightarrow B(k)+\alpha g[B(t)],
\end{equation}
where $\alpha$ denotes the learning rate, subject to variation based on the number of iterations $k$.

Guided by Eq.~(\ref{Eq_time}), the determination of $B(t)$ involves considering two forms of guess fields. One takes on a Gaussian profile, defined as
\begin{equation}
B_{\textrm{G}}=\frac{B_0}{\sqrt{2\pi}\sigma}e^{-\frac{t_{g}^2}{2\sigma^2}},
\end{equation}
with $B_0=J$, $\sigma=0.1$, and $t_{g}=[-0.5,0.5]$, while the other manifests as a random sequence comprising random numbers $\xi$ within the range $[0,1]$ noted as $B_{\rm G}=B_{0}\xi$. The guess fields in the GRAPE optimal algorithm consist of discrete points with a time interval of $\delta t=T/n$, where $n$ takes values of $\{100,10\}$ for Gaussian and random types, respectively. Fig.~\ref{Multi_Per} illustrates the optimized results for $N={3\sim6}$. Within the GRAPE optimal algorithm, the maximum population of the target state is achievable after optimization, irrespective of the shape of the guess field, once the evolution time is specified. The corresponding evolution time $T$ and the optimized population $P(N)_{O}$ for varying particle numbers are presented in Table~\ref{Tb_Per}.
\begin{table}
\centering
\caption{\label{Tb_Per}The evolution time and the populations of the multiparticle complete graph states governed by Eq.~(\ref{Eq_time}).}
\setlength{\tabcolsep}{5.5mm}
\begin{tabular}{ccc}
\hline\hline
Target state&Evolution time (units of $J$)&Population \\
$|K_{3}\rangle$&2.3&1 \\
$|K_{4}\rangle$&2.808&0.9931 \\
$|K_{5}\rangle$&3.386&0.9710 \\
$|K_{6}\rangle$&3.952&0.9346 \\
\hline\hline
\end{tabular}
\end{table}
The results demonstrate the successful generation of multiparticle complete graph states under the influence of a spin-$1/2$ Heisenberg $XX$ chain in a time-varying magnetic field optimized by the GRAPE algorithm. Consequently, the preparation of multiparticle complete graph states can be achieved mathematically in a ``one-step" manner through Eq.~(\ref{Eq_time}). This protocol is universally applicable in systems capable of nearest-neighbor interactions characteristic of the spin-$1/2$ Heisenberg $XX$ chain, including superconducting quantum circuits \cite{PhysRevA.91.022315,Roushan2017,science.aao1401,Yanay2020} and ion traps \cite{Graß2014,science.aau4963,PhysRevResearch.2.023015,PhysRevB.105.L241103,PhysRevX.13.031017}.
In the following part, we will delve into the detailed methodology for preparing the aforementioned multiparticle complete graph states within a neutral-atom system, closely considering experimental parameters.

\section{A specific implementation based on Rydberg atom array}\label{IV}
\begin{figure*}
\centering\scalebox{0.3}{\includegraphics{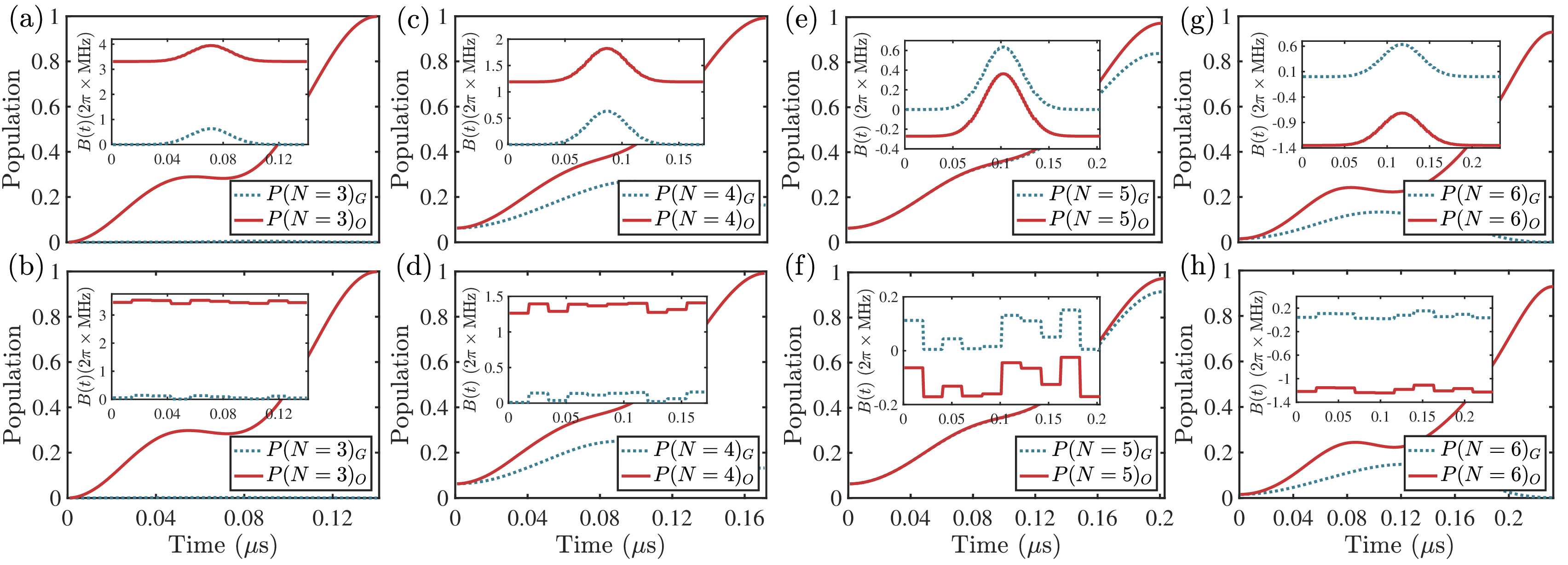}}
\caption{\label{Multi_vdW}The realization of the multiparticle complete graph states with $N=3\sim6$ governed by the Hamiltonian $H_{\textrm{sys}}$ initialized from state $|\Psi_{0}\rangle=\bigotimes_{i=1}^{N}[{1}/{\sqrt{2}}(|\uparrow\rangle+|\downarrow\rangle)]$. (a) and (b), respectively, show evolution under the magnetic field in Gaussian and random types for $N=3$. The dotted lines correspond to the evolution under the guess field, whereas the solid lines correspond to the optimized one. Then (c) and (d) show the evolution for $N=4$, while (e) and (f) for $N=5$, (g) and (h) for $N=6$. The insets show the corresponding time-dependent $B(t)$ of the initial guess and the optimized result, where $B_{0}=2\pi\times1~{\rm MHz}$.}
\end{figure*}

The Rydberg atom array, characterized by robust programmability, has recently emerged as a powerful candidate in quantum simulations, spurred by breakthroughs in experimental technology \cite{Pause:24,PhysRevLett.130.180601}. Individual Rydberg atoms, for example, can be placed in an optical tweezers array \cite{science.aah3752,Barredo2018}. Leveraging the long-range interactions inherent in highly excited Rydberg states, the Rydberg atom array finds widespread applications in quantum state transmission \cite{PhysRevA.95.013403}, quantum topology research \cite{science.aav9105,Scholl2021,PhysRevLett.129.090401}, and artificial gauge field \cite{PhysRevLett.112.043001,PhysRevA.106.L021101}. Notably, recent works, such as those in Refs.~\cite{Ebadi2021,Bluvstein2022}, have realized a primary quantum computer employing 256 Rydberg atoms, demonstrating proficiency in various quantum tasks.

Within the Rydberg atom array, the Rydberg dipole-dipole interaction serves as a direct means to achieve $H_{\textrm{XX}}$ \cite{Bornet2023,chen2023spectroscopy}. Additionally, the magnetic field $B(t)$ can be equivalently obtained through the Stark effect induced by a large detuned time-dependent laser field coupled transition between the intermediate state $|p\rangle$ and one of the Rydberg states \cite{Chen2023,bornet2024enhancing}. It is crucial to note that the inevitable long-range interactions among Rydberg atoms, such as van der Waals (vdW) forces and the long-range dipolar interactions, should also be taken into consideration, with respect to the realistic physical system.

\subsection{The realization of multiparticle complete graph states}

We consider a one-dimensional chain geometry with open periodic boundary conditions, as shown in Fig.~\ref{Complete}(b),
in which $N$ atoms are contained in tweezers that are equidistantly spaced $R$ along the quantized $z$ axis.
For the purpose of illustration, states with opposite parity are transformed into pseudo-spin states, using the $^{87}\textrm{Rb}$ atom as an example, $|\uparrow\rangle=|80S_{1/2},m_{j}=1/2\rangle$ and $|\downarrow\rangle=|79P_{3/2},m_{j}=1/2\rangle$ \cite{Browaeys_2016}.
The resonant dipole-dipole interaction gives rise to the Heisenberg $XX$ Hamiltonian
\begin{equation}
H_{\textrm{XX}}=\sum_{\langle i,j\rangle}\frac{V_{\textrm{dip}}(\theta)}{2}(\sigma^{x}_{i}\sigma^{x}_{j}+\sigma^{y}_{i}\sigma^{y}_{j}),
\end{equation}
where $V_{\textrm{dip}}(\theta)=C_{3}(1-3\cos^2\theta)/R^3$ with $\theta$ the polar angle between the quantization $z$ axis and the vector direction of atomic connection, and the constant $C_{3}$ is defined as
$2\pi\times8.780~{\rm GHz}\cdot\mu{\rm m}^3$ \cite{SIBALIC2017319}.
For $\theta=0$ and $R=19.3~\mu{\rm m}$, we have $V_{\textrm{dip}}\simeq-2\pi\times2.443~{\rm MHz}$.
Additionally, we account for inevitable interactions, including long-range vdW interactions and dipolar interactions, which are characterized as error terms.
\begin{eqnarray}\label{Eq_err}
H_{\textrm{err}}&=&\sum_{j>i}(U^{\uparrow}_{ij}|\uparrow_{i}\uparrow_{j}\rangle\langle\uparrow_{i}\uparrow_{j}|+U^{\downarrow}_{ij}|\downarrow_{i}\downarrow_{j}\rangle\langle\downarrow_{i}\downarrow_{j}|)\nonumber\\&&+\sum_{j>i+1}\frac{V_{\textrm{dip}}^{ij}}{2}(\sigma^{x}_{i}\sigma^{x}_{j}+\sigma^{y}_{i}\sigma^{y}_{j}),
\end{eqnarray}
where $U^{\uparrow(\downarrow)}_{ij}=-C^{\uparrow(\downarrow)}_{6}/R_{ij}^6$ with $C^{\uparrow}_{6}=-2\pi\times4161.55~{\rm GHz}\cdot\mu{\rm m}^6$ and $C^{\downarrow}_{6}=2\pi\times3452.60~{\rm GHz}\cdot\mu{\rm m}^6$, and $V_{\rm dip}^{ij}=C_{3}(1-3\cos^2\theta)/R_{ij}^3$. $R_{ij}$ denotes the distance between the $i$th and $j$th atom.
\begin{equation}\label{system}
H_{\rm sys}=H_{\rm XX}+H_{\rm err}+\sum_{i=1}^{N}B(t)S_{i}^{z}.
\end{equation}

Using $N=3\sim6$ as illustrative examples, we depict the evolutions governed by Eq.~(\ref{system}). In Fig.~\ref{Multi_vdW}, we specifically show the evolutions under the initial guess and optimized fields in Gaussian and random types. Within this framework, the populations in the target state, denoted as $P(N)_{O}$, exhibit multiple peak points corresponding to different values of $T$ (as shown in the picture in Appendix~\ref{app1}). By scanning the time within $0.25~\mu{\rm s}$, we successfully obtain the optimized fields under $H_{\textrm{sys}}$ for various numbers of particles. The inset shows the time-dependent $B(t)$ of the application, where solid lines represent the optimized pulse, and dotted lines represent the guess pulse. The evolution time $T$ and the populations of the target states are summarized in Table~\ref{multi}. Although the population of the target state decreases as $N$ increases, we can still prepare the six-particle complete state with a population greater than $0.92$.

\begin{table}
\centering
\caption{\label{multi}The evolution time and the populations of the multiparticle complete graph states corresponding to the system constructed by $^{87}{\rm Rb}$ atom array arranged in a line with polar angle $\theta=0$.}
\setlength{\tabcolsep}{5.5mm}
\begin{tabular}{ccc}
\hline\hline
Target state&Evolution time ($\mu{\rm s}$)&Population \\
$|K_{3}\rangle$&0.141&0.9989 \\
$|K_{4}\rangle$&0.172&0.9920 \\
$|K_{5}\rangle$&0.203&0.9728 \\
$|K_{6}\rangle$&0.233&0.9294 \\
\hline\hline
\end{tabular}
\end{table}

Recently, a quantum circuit utilizing the Rydberg atom array has been demonstrated to generate a complete graph state of five particles \cite{crescimanna2023quantum}. When coupled with quantum annealing algorithms, the fidelity of the target state can be increased from approximately $0.85$ to $0.999$ by increasing computational depth. However, this increased computational depth is accompanied by prolonged execution times, leading to an increased susceptibility to decoherence effects. On the contrary, our protocol enables the generation of the target state in a single step, regardless of the number of atoms involved. This approach significantly reduces operational time while preserving high levels of fidelity, which presents notable advantages in mitigating decoherence effects.

\subsection{Experimental feasibility analysis}

This part focuses on experimental feasibility by examining the experimental imperfections and providing a concise design for an experiment procedure using the Rydberg atom array. This includes the production of the initial state and the final decoupling process.

\subsubsection{The influence of experimental errors}

Here, we rigorously assess the scheme by taking into account experimental imperfections, such as spontaneous radiation of Rydberg levels, as well as the discrepancy between the dipole-dipole interaction $V_{\textrm{dip}}$ and the magnetic field $B(t)$. The associated numerical simulations employ Gaussian-type optimal pulses.

(i) Examining the impact of spontaneous emission of Rydberg states. To analyze the influence of spontaneous radiation, we introduce an empty ground state $|g\rangle$,
under the assumption that all Rydberg states will spontaneously radiate to this unoccupied state. At a temperature of $0.1~\rm K$, the lifetimes of the Rydberg states $|80S_{1/2}\rangle$ and $|79P_{3/2}\rangle$ are $\tau_{\uparrow,\downarrow}=0.569,1.1~{\rm ms}$, respectively. Thus, the corresponding master equation can be described as
\begin{eqnarray}
\mathcal{L}[\rho]&=&-i[H_{\rm sys},\rho]+\sum_{i=1,j=\uparrow,\downarrow}^{N}\frac{\gamma_{j}}{2}(2s_{i}\rho s_{i}^{+}-s_{i}^{+}s_{i}\rho\nonumber\\&&-\rho s_{i}^{+}s_{i}),
\end{eqnarray}
where $\gamma_{\uparrow(\downarrow)}$ is the decay rate of the Rydberg state $|\uparrow(\downarrow)\rangle$ equals to $1/\tau_{\uparrow(\downarrow)}$, and $s_{i}=|g\rangle_{i}\langle j|$ denotes the decay channel of the $i$th atom.
Through numerical simulation, dissipation is observed to lead to a reduction in the population of the target state by values of $\{0.0006,0.0009,0.0010,0.0017\}$, corresponding to
$N$ ranging from three to six, respectively.

(ii) Addressing the impact of the mismatch between the dipole-dipole interaction $V_{\textrm{dip}}$ and the magnetic field $B(t)$. This discrepancy may arise from sources: the instability in the distance between atoms and the fluctuation of the magnetic field.

\begin{figure}
\centering\scalebox{0.32}{\includegraphics{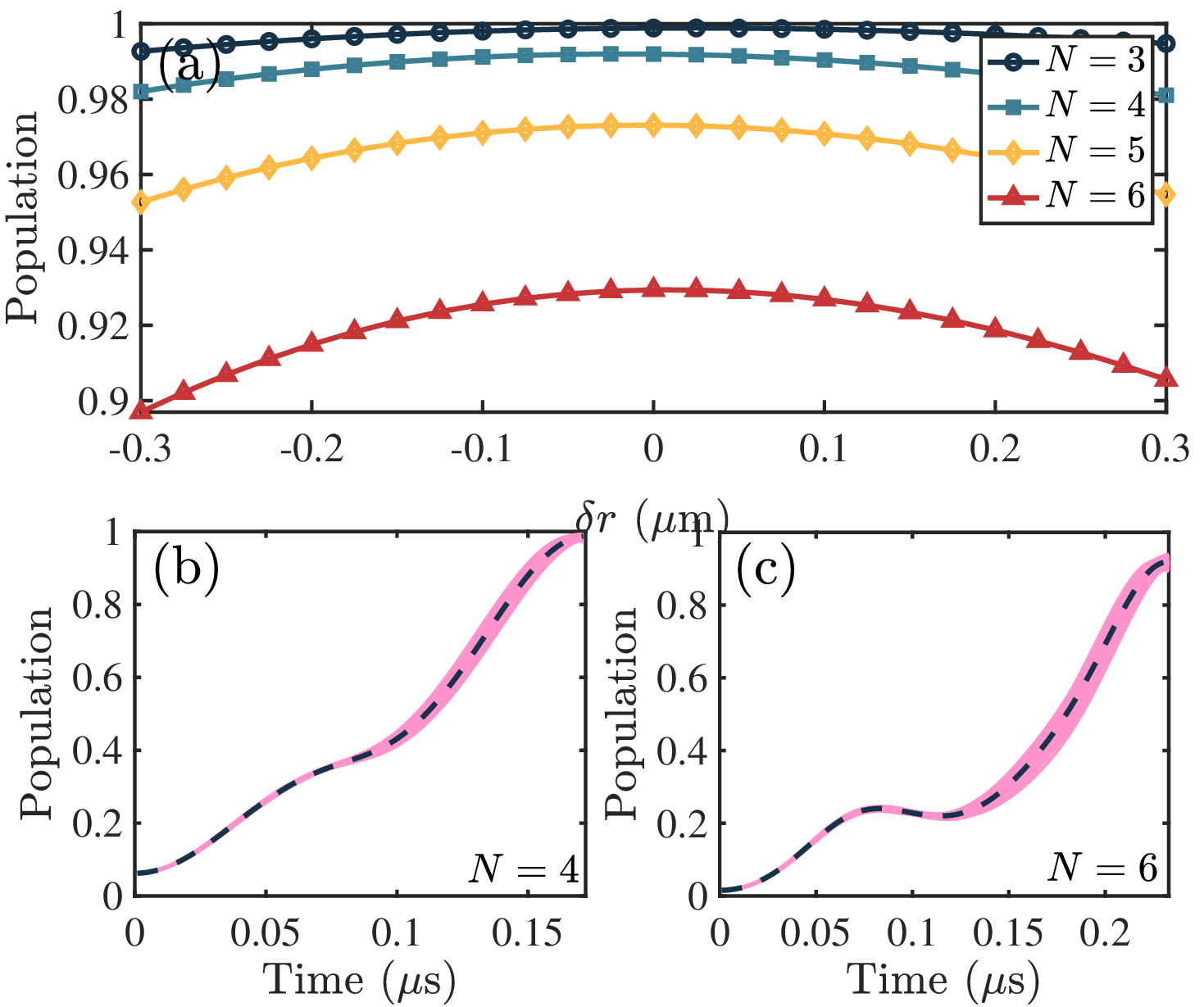}}
\caption{\label{Mismatch_dU}The influence of the mismatch of the dipole interaction $V_{\textrm{dip}}$ and the magnetic field $B(t)$ caused by atomic vibration. (a) Populations of the complete multiparticle graph states with range error $\delta r$ under the Hamiltonian $H_{\textrm{sys}}(\delta r)$. (b) and (c) Average evolution results for $N=4,6$ considering a true fluctuation where the position distribution is Gaussian with variance $\{\sigma_{x},\sigma_{y},\sigma_{z}\}\simeq\{193.5,193.5,1242.9\}~{\rm nm}$. Note that the light-red region shows the results of 50 stochastic simulations, and the dashed lines correspond to the average results.
}
\end{figure}
\begin{figure}
\centering\scalebox{0.32}{\includegraphics{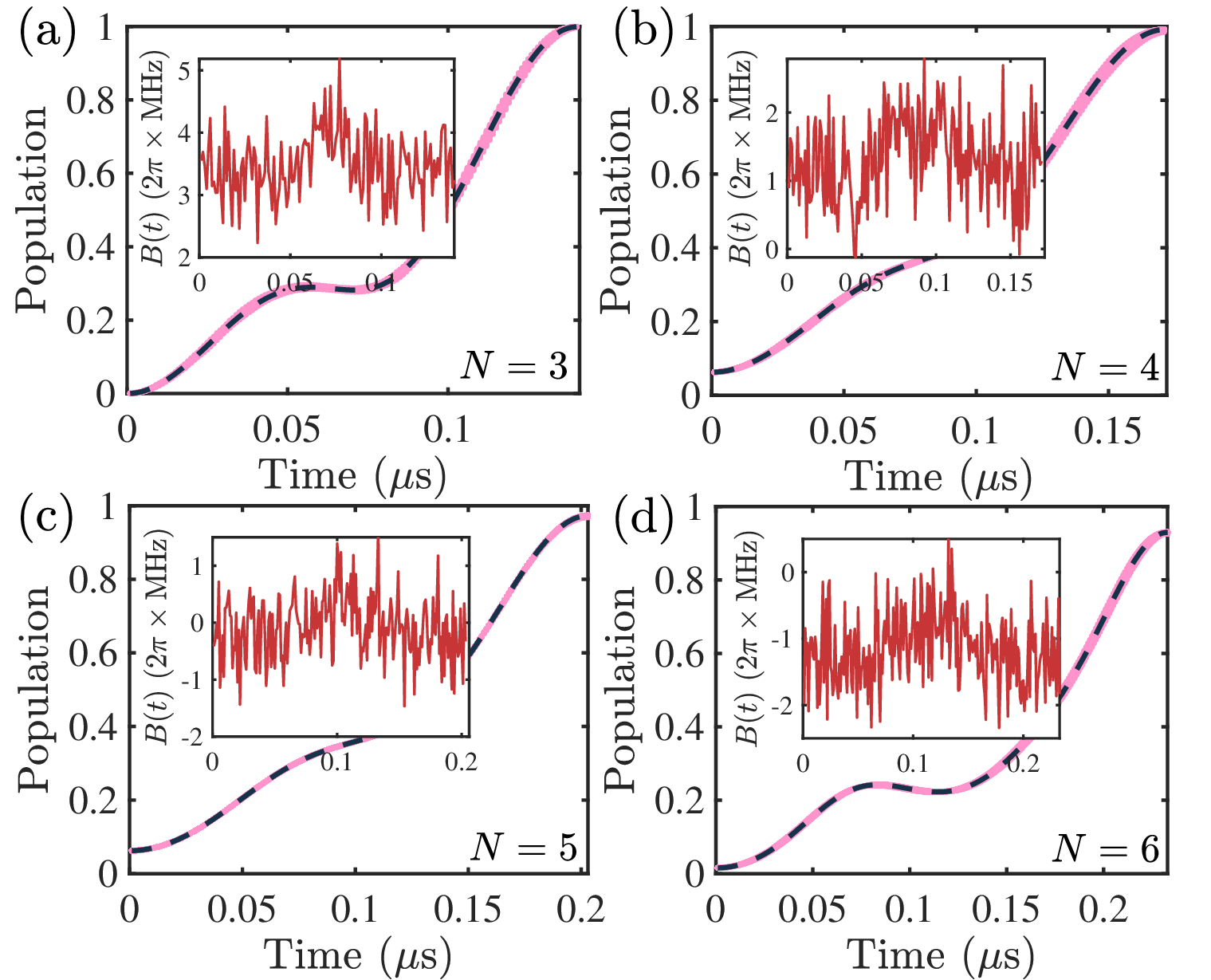}}
\caption{\label{Mismatch_dM}The influence of the mismatch of $V_{\textrm{dip}}$ and $B(t)$ caused by the fluctuation of the magnetic field. (a)-(d), respectively, show the average evolution results (dashed line in blue) governed by Eq.~(\ref{Eq_dM}). Note that the light-red region shows the results of 50 stochastic simulations, while the inner pictures present the magnetic field of one of the experiments.}
\end{figure}

To investigate the first, we introduce $V_{\textrm{dip}}(\delta r)=-2C_{3}/(R+\delta r)^{3}$ with a range error $\delta r\in[-300,300]~{\rm nm}$ for a preliminary estimate. The Hamiltonian $H(t)$ can be rewritten as
\begin{equation}\label{Eq_dU}
H_{\delta V}(t)=\sum_{\langle i,j\rangle}\frac{V_{\textrm{dip}}(\delta r)}{2}(\sigma^{x}_{i}\sigma^{x}_{j}+\sigma^{y}_{i}\sigma^{y}_{j})+\sum_{i=1}^{N}B(t)S^{z}_{i}.
\end{equation}
It should be noted that as the vdW interaction also depends on the distance between the atoms, the Hamiltonian $H_{\textrm{err}}$
should likewise be formulated as
\begin{eqnarray}\label{Eq_err}
H^{\textrm{err}}_{\delta V}&=&\sum_{j>i}[U^{\uparrow}_{ij}(\delta r)|\uparrow_{i}\uparrow_{j}\rangle\langle\uparrow_{i}\uparrow_{j}|
+U^{\downarrow}_{ij}(\delta r)|\downarrow_{i}\downarrow_{j}\rangle\langle\downarrow_{i}\downarrow_{j}|]\nonumber\\&&
+\sum_{j>i+1}\frac{V_{\textrm{dip}}^{ij}(\delta r)}{2}(\sigma^{x}_{i}\sigma^{x}_{j}+\sigma^{y}_{i}\sigma^{y}_{j}),
\end{eqnarray}
where $U^{\uparrow(\downarrow)}(\delta r)=-C^{\uparrow(\downarrow)}_{6}/(R_{ij}+\delta r)^6$, and $V_{\rm dip}^{ij}=C_{3}(1-3\cos^2\theta)/(R_{ij}+\delta r)^3$.
In Fig.~\ref{Mismatch_dU}(a), we present the evolution results governed by $H_{\textrm{sys}}(\delta r)=H_{\delta V}(t)+H^{\textrm{err}}_{\delta V}$, with $N$ varying from three to six, respectively. The population of target states $P(N)_{O}$ consistently exceeds 0.9 for $\delta r \leq 300~{\rm nm}$. In experimental settings, the vibration of atoms near the ideal position, induced by nonzero temperature, follows the Maxwell-Boltzmann distribution and is associated with the atomic temperature as well as the parameters of the tweezer beam and trap. According to Ref.~\cite{PhysRevA.105.042430}, the atomic temperature can be cooled to $5.2~\mu{\rm K}$ in a $50~\mu{\rm K}$ trap. Setting the corresponding parameters of the laser beams as wavelength $\lambda_{f}=830~{\rm nm}$, typical beam power $P_{f}=174~\mu{\rm W}$, and waist ($1/e^2$ intensity radius) $\omega_{f}=1.2~\mu{\rm m}$, we estimate a position distribution in Gaussian with variance $\{\sigma_{x},\sigma_{y},\sigma_{z}\}\simeq\{193.5,193.5,1242.9\}~{\rm nm}$.
Thus, in Fig.~\ref{Mismatch_dU}(b) and Fig.~\ref{Mismatch_dU}(c), we further assess the robustness of the protocol with $N=4$ and $N=6$ as realistically as possible, considering a three-dimensional random vibration under these estimated parameters. The distance between the atoms $R$ is redefined as $D=|\textbf{R}_{i}-\textbf{R}_{i+1}|$, where $\textbf{R}_{i}=(x_{i},y_{i},z_{i})$. Since the evolution time $T$ is less than $0.25~\mu{\rm s}$, we reasonably consider only one group of random fluctuations ($\sigma_{x,y,z}$) error throughout the evolution process here. Fig.~\ref{Mismatch_dU}(b) and \ref{Mismatch_dU}(c) respectively illustrate the population of the target states $|K_{4}\rangle$ and $|K_{6}\rangle$, where the light red portions represent the results of 50 stochastic simulations, and the dashed lines in dark blue correspond to the average outcomes. The average populations of $|K_{4}\rangle$ and $|K_{6}\rangle$ in this test are $0.9728$ and $0.9187$, respectively. With advancements in experimental methods, the actual vibration of the atoms in the trap is expected to be smaller than our estimated parameters \cite{Labuhn2016,Chew2022}, which is more conducive to the realization of our scheme.

For the latter reason, the fluctuation of the magnetic field intensity is introduced by a time-dependent fluctuation $\delta B$. It is assumed to follow normal distribution functions with standard deviations $\sigma(t)\simeq2\pi\times0.5~{\rm MHz}$. Thus, we have the system Hamiltonian
\begin{equation}\label{Eq_dM}
H_{\delta B}(t)=H_{\rm XX}+H_{\rm err}+\sum_{i=1}^{N}[B(t)+\delta B]S^{z}_{i}.
\end{equation}
As shown in Fig.~\ref{Mismatch_dM}, we present the evolution results governed by the Hamiltonian $H_{\delta B}(t)$ averaged over 50 realizations. It is evident that the fluctuation in the global magnetic field has a negligible impact on our scheme.

\subsubsection{Mapping between Rydberg states and ground states}

\begin{figure}
\centering\scalebox{0.21}{\includegraphics{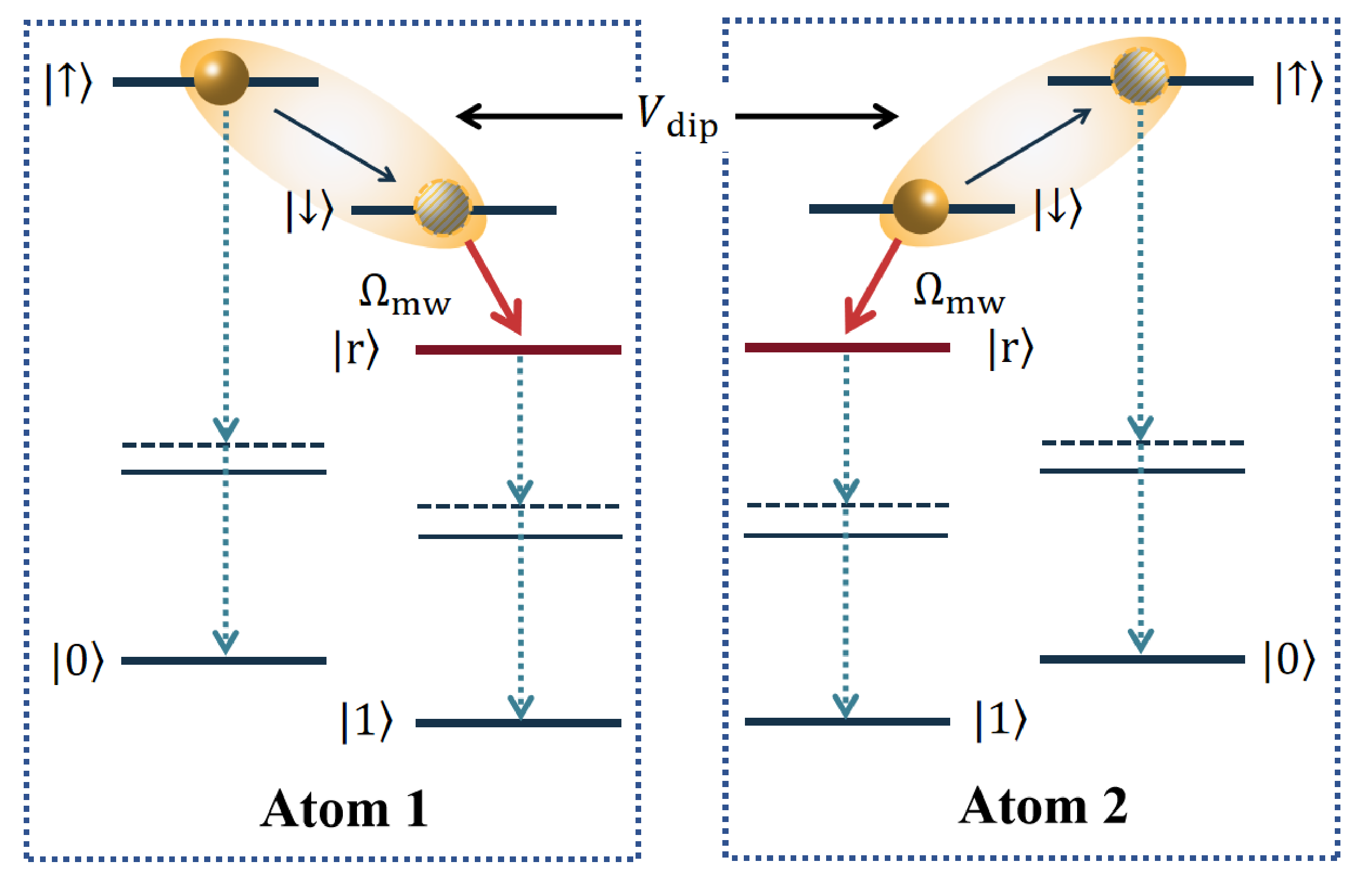}}
\caption{\label{Decoupling}The decoupling process (dashed lines) and the relevant levels of $^{87}{\rm Rb}$ atom. The ground states can be chosen as $|5S_{1/2}\rangle$ hyperfine clock states, such as $|0\rangle\equiv|F=1, m_{F}=0\rangle$, $|1\rangle\equiv|F=2, m_{F}=0\rangle$, while the auxiliary Rydberg states can be chosen as  $|r\rangle=|78S_{1/2},m_{j}=1/2\rangle$.}
\end{figure}

For the neutral-atom system, ensuring the stability of the target state typically involves encoding it in the ground states $|0\rangle$ and $|1\rangle$, which are the hyperfine clock states of $|5S_{1/2}\rangle$. Moreover, given the natural existence of dipole-dipole interactions between Rydberg states, the scheme should be implemented in three steps, that is, the initial state preparation, the core evolution governed by $H_{\textrm{sys}}$, and the decoupling process. In this section, we present separate feasibility proposals for the initial state preparation and decoupling process.

(i) Preparation of the initial state. In this protocol, the initial state is selected as
\begin{equation}
|\Psi_{N}(0)\rangle=\bigotimes_{i=1}^{3}[\frac{1}{\sqrt{2}}(|\uparrow_i\rangle+|\downarrow_i\rangle)].
\end{equation}
To excite multiple atoms to Rydberg states without being constrained by the Rydberg blockade, the process can be initiated by driving the ground state $|0\rangle\rightarrow|\uparrow\rangle$ through a two-photon resonance with an effective Rabi frequency $\Omega$ and an evolution time $t=\pi/\Omega$. Subsequently, the transition from $|\uparrow\rangle$ to $|\downarrow\rangle$ is achieved sequentially using a microwave pulse with Rabi frequency $\Omega_{\textrm{mw}}^{a}$ and duration $t_{\textrm{mw}}=\pi/(2\Omega_{\textrm{mw}}^{a})$. This technique to prepare the initial state is well established in quantum simulations \cite{PhysRevLett.120.063601,PRXQuantum.3.020303,science.aav9105}.

(ii) Dynamical decoupling process: As dipole-dipole interactions naturally persist, the system evolution under the Hamiltonian $H_{\textrm{XX}}$ cannot be voluntarily stopped by controlling external electric or magnetic fields. To address this challenge, we select an auxiliary Rydberg level $|r\rangle$ characterized by non-dipole-dipole interactions,
such as $|78S_{1/2},m_{j}=1/2\rangle$. Illustrated by the dashed line in Fig.~\ref{Decoupling}, the dynamics can be decoupled by driving atoms in states $|\downarrow\rangle$ to this auxiliary Rydberg level via a microwave pulse $\Omega_{\textrm{mw}}^{b}$. Given that the Rabi frequency of the microwave pulse can reach $200~{\rm MHz}$ \cite{science.abd9547}, this process is rapid. By combining this step with the two-photon resonance process with an effective Rabi frequency $\Omega$, we can further drive the Rydberg levels to hyperfine clock states with an evolution time $t=\pi/\Omega$, thus obtaining complete graph states encoded by ground states of the $^{87}{\rm Rb}$ atom.

\begin{figure}
\centering\scalebox{0.32}{\includegraphics{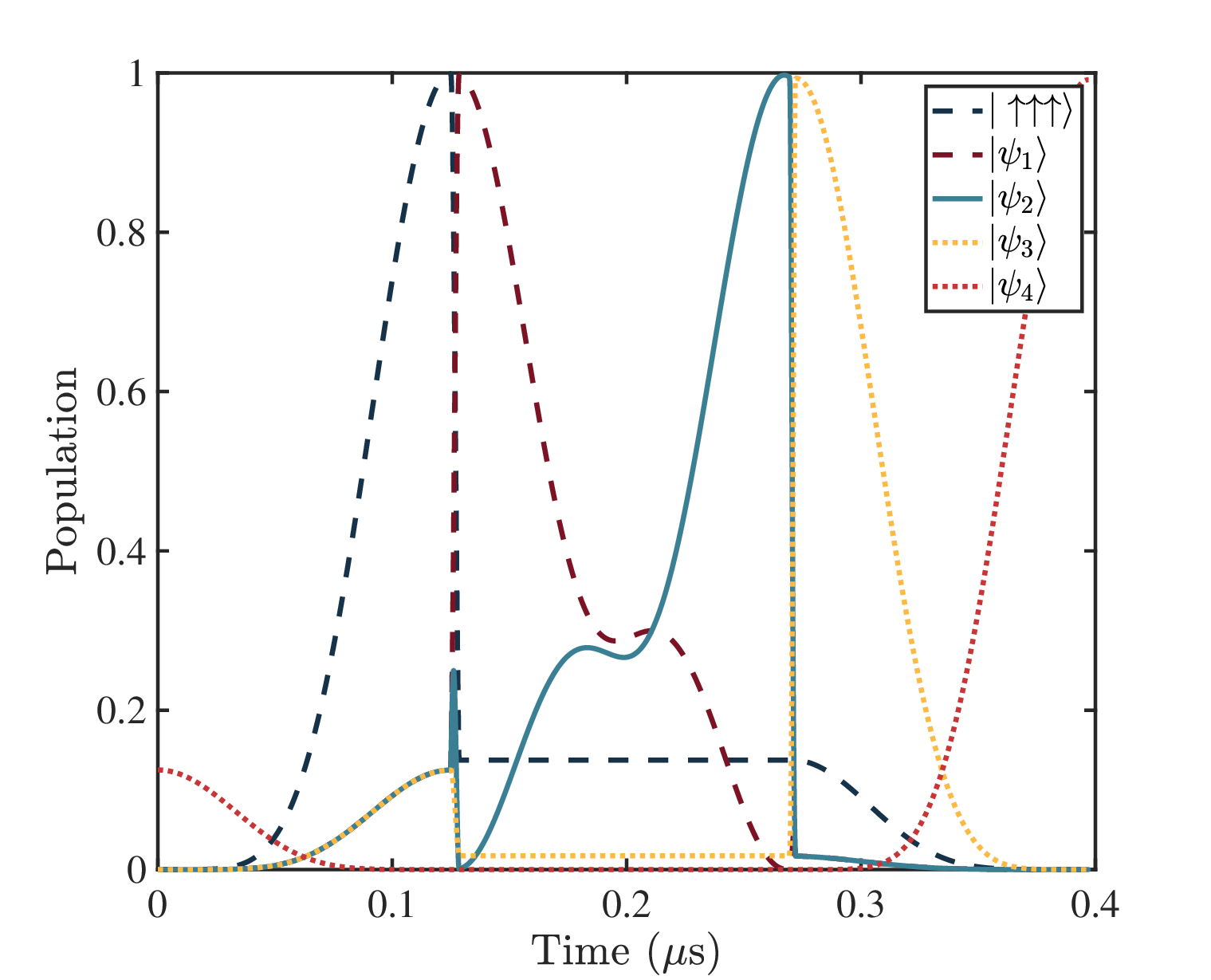}}\caption{\label{Three}A complete evolution via the resonant dipole-dipole interaction for $N=3$ with initial state $|000\rangle$. The dashed and dotted lines respectively correspond to the initial state preparation and the decoupling process, and the solid line presents the core evolution with the optimized $B(t)$ in Gaussian type. The parameters are set as $\Omega=2\pi\times4~{\rm MHz}$, $\Omega_{\rm mw}^{a}=2\pi\times70~{\rm MHz}$, and $\Omega_{\textrm{mw}}^{b}=2\pi\times200~{\rm MHz}$.}
\end{figure}

Fig.~\ref{Three} illustrates a complete evolution with $N=3$. To enhance observability, we depict the evolution of the target states for each process, where $|\psi_1\rangle=\bigotimes_{i=1}^{3}[{1}/{\sqrt{2}}(|\uparrow_{i}\rangle-i|\downarrow_{i}\rangle)]$, $|\psi_{2}\rangle={1}/{(2\sqrt{2})}\bigotimes_{i=1}^{3}[|\uparrow_{i}\rangle(-1)^{3-i}\prod_{j=i+1}^{3}\sigma_{z}^{j}-i|\downarrow_{i}\rangle]$, $|\psi_{3}\rangle={1}/{(2\sqrt{2})}\bigotimes_{i=1}^{3}[|\uparrow_{i}\rangle(-1)^{3-i}\prod_{j=i+1}\sigma_{z}^{j}-|r_{i}\rangle]$, and $|\psi_{4}\rangle={1}/{(2\sqrt{2})}\bigotimes_{i=1}^{3}[-|0_{i}\rangle(-1)^{3-i}\prod_{j=i+1}\sigma_{z}^{j}+|1_{i}\rangle]$. The dashed and dotted lines correspond to the initial state preparation and the decoupling process, respectively, while the solid line represents the core evolution with the optimized $B(t)$ in Gaussian type. Under the parameters $\Omega=2\pi\times4~{\rm MHz}$, $\Omega_{\rm mw}^{a}=2\pi\times70~{\rm MHz}$ and $\Omega_{\textrm{mw}}^{b}=2\pi\times200~{\rm MHz}$, the population of the target state can reach about $0.9916$ with a total evolution time $t_{\textrm{tot}}=0.3971~\mu{\rm s}$. Given that the independent evolution process of the core operation can reach perfection with $N=3$, the error caused by the preparation of the initial state and the return driving is approximately $0.0074$.

It should be noted that this three-step scheme is specifically designed for the Rydberg atom array system. In other systems, such as superconducting systems, the mapping process mentioned here may no longer be necessary.

\begin{table}
\centering
\caption{\label{err}Error estimation of population for the multiparticle complete graph states with $N=3\sim6$ corresponding to the system constructed by Rydberg atom array.}
\setlength{\tabcolsep}{1.2mm}
\begin{tabular}{ccccc}
\hline\hline
Error budget&$|K_{3}\rangle$&$|K_{4}\rangle$&$|K_{5}\rangle$&$|K_{6}\rangle$ \\
Initial preparing and decoupling&\multicolumn{4}{c}{0.0074} \\
Spontaneous radiation&0.0006&0.0009&0.0010&0.0017\\
Atomic vibration ($\pm100~{\rm n m}$)&0.0009&0.0016&0.0020&0.0038\\
\hline
$P(N)_{O}$&0.99&0.9821&0.9624&0.9165 \\
\hline\hline
\end{tabular}
\end{table}
As shown in Table.~\ref{err}, we make an estimation on the population of complete states with $N=3\sim6$ considering the experimental errors to make a comparison with the other protocol. In recent experiments, the fidelity of the two-qubit entanglement gate has been improved to $99.5\%$ \cite{Jandura2022timeoptimaltwothree,Evered2023}. Based on this analysis, we can estimate the time and fidelity necessary to achieve complete graph states using CZ gates for comparison. For $N=3$, a minimum of three CZ gates are required, with an estimated fidelity of approximately $0.985$. For $N=6$, 15 CZ gates are needed, with an estimated fidelity of $0.928$, demonstrating performance comparable to our approach. However, for the gate-based approach, although a single gate operation time can be less than $300~{\rm ns}$, such CZ gate operations cannot be performed simultaneously. The evolution time will increase rapidly with increasing number of particles $N$. On the contrary, the evolution time required by our protocol is shorter, which can avoid some unknown errors. This reduction is crucial from the perspective of decoherence.

\section{summary}\label{V}
In summary, we have introduced an efficient single-step technique for generating multiparticle complete graph states. Our approach relies solely on a constant nearest-neighbor interaction strength and a time-varying magnetic field which is optimized using the GRAPE algorithm. This methodology is applicable to various physical systems capable of implementing a spin-$1/2$ Heisenberg $XX$ chain, such as superconducting systems, trapped ion systems, and neutral atom systems.
We have thoroughly investigated the protocol's execution in the Rydberg atom array system, evaluating its resilience in the face of experimental errors. Our numerical findings underscore the robustness of our approach against challenges such as atomic position inaccuracies and pulse oscillations.
Our one-step solution significantly reduces operating time while maintaining a relatively high estimated fidelity. We anticipate that this discovery will quickly contribute to the practical implementation of quantum computing and quantum error correction in neutral-atom systems.
\begin{figure}
\centering\scalebox{0.32}{\includegraphics{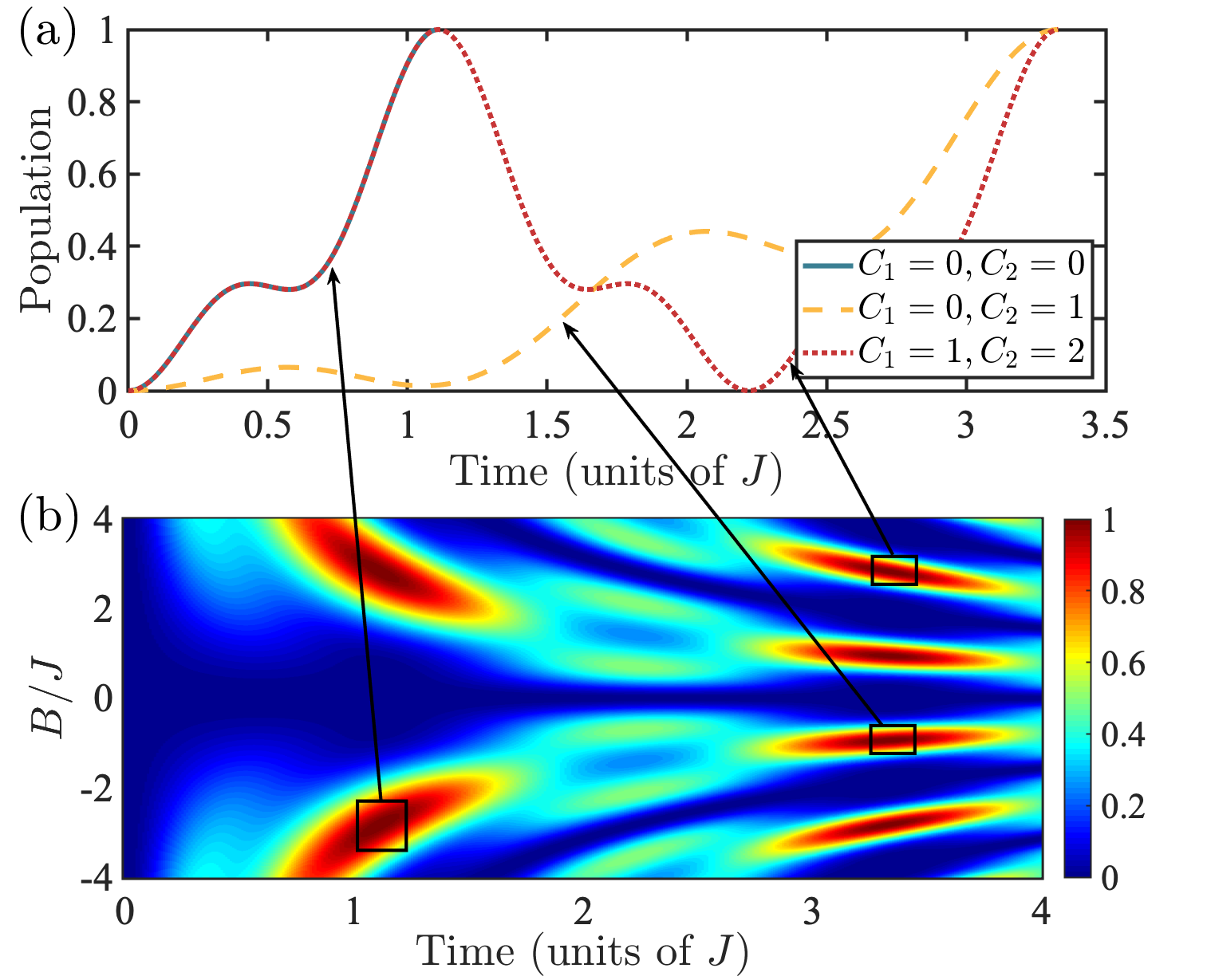}}\caption{\label{Constent}Evolution governed by the Hamiltonian (\ref{Eq_con}) with $N=3$. (a) The intensity of the constant amplitude pulse $B$ satisfies the relationship shown in Eq.~(\ref{Eq_rela}). (b) Scanning results for the population of the target state with different parameters.}
\end{figure}

\begin{figure}
\centering\scalebox{0.32}{\includegraphics{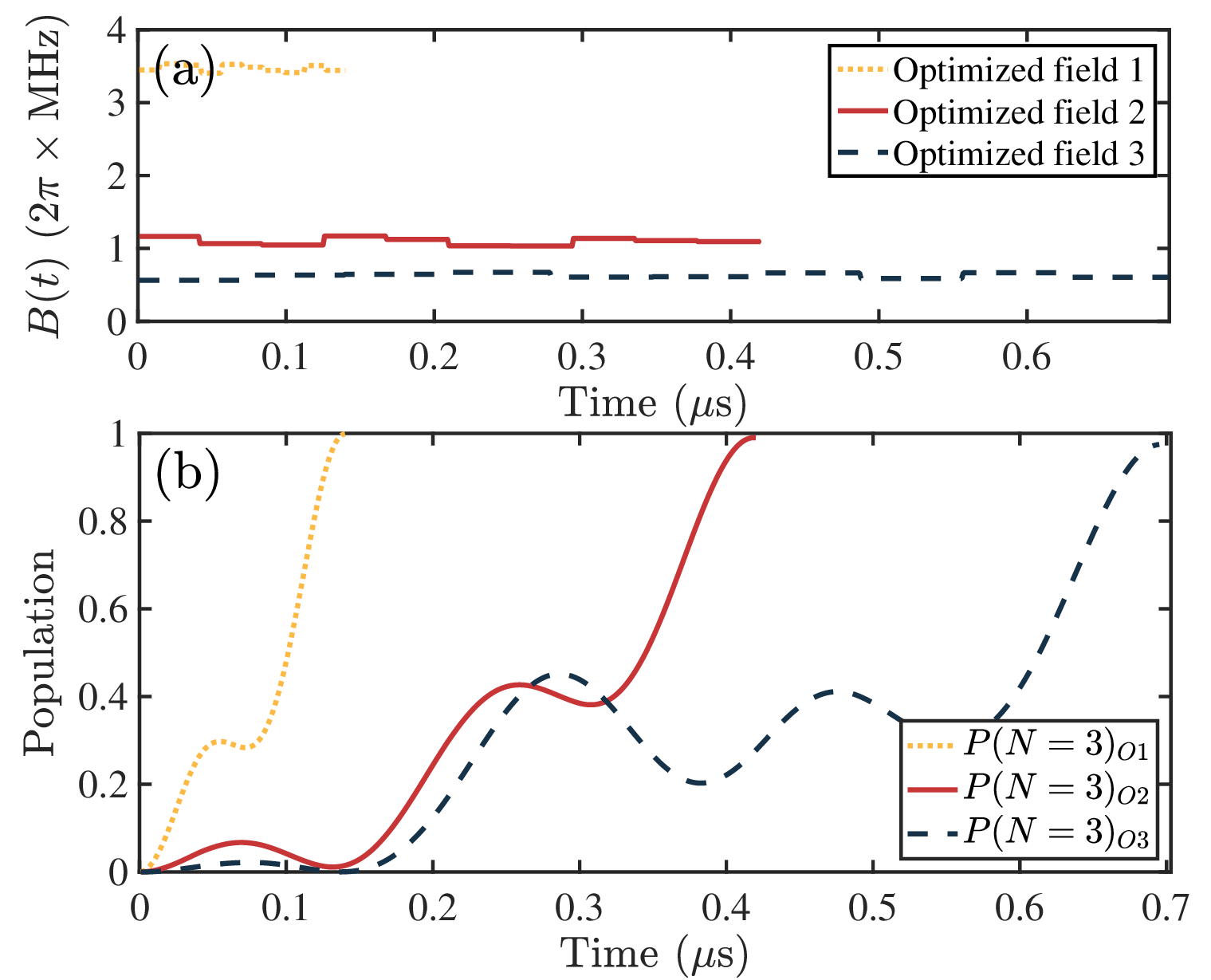}}\caption{\label{Multi_T}The multiple peaks corresponding to different $T$. (a) The optimized magnetic fields $B(t)$ under different evolution time $T$. (b) The corresponding evolution governed by $H_{\rm sys}$ with $N=3$.}
\end{figure}

\section*{acknowledgment}
The authors thank Professor Re-Bing Wu of Tsinghua University for his guidance on the GRAPE optimal algorithm. The anonymous reviewers are also thanked for constructive
comments that helped improve the quality of this article. This work is supported by the NFSC through Grant No. 12174048, the Fundamental Research Funds for the Central Universities (Grant No. 3122023QD26) and the Research Start-up Funds of the Shenyang Normal University (Grants No. BS202301, No. BS202303, No. BS202314).

\appendix
\section{A protocol for $N=3$ under constant amplitude magnetic field}\label{app}
In the three-particle case, the preparation of the complete graph state is particularly unique. Governed by the Hamiltonian with a time-independent global magnetic field, represented as
\begin{equation}\label{Eq_con}
H_{\textrm{con}}=\sum_{\langle i,j\rangle}J_{ij}(\sigma^{x}_{i}\sigma^{x}_{j}+\sigma^{y}_{i}\sigma^{y}_{j})+\sum_{i=1}^{3}B S^{z}_{i},
\end{equation}
the system has definite solutions to realize the corresponding complete state.
According to Schr\"{o}dinger equation,
\begin{equation}
i\frac{d|\Psi(t)\rangle}{dt}=H|\Psi(t)\rangle,
\end{equation}
we have
\begin{eqnarray}
|\Psi(t)\rangle&=&\psi_{1}(t)|\downarrow\downarrow\downarrow\rangle+
\psi_{2}(t)|\downarrow\downarrow\uparrow\rangle+\psi_{3}(t)|\downarrow\uparrow\downarrow\rangle\nonumber\\&&+
\psi_{4}(t)|\downarrow\uparrow\uparrow\rangle+\psi_{5}(t)|\uparrow\downarrow\downarrow\rangle+
\psi_{6}(t)|\uparrow\downarrow\uparrow\rangle\nonumber\\&&+\psi_{7}(t)|\uparrow\uparrow\downarrow\rangle+
\psi_{8}(t)|\uparrow\uparrow\uparrow\rangle,
\end{eqnarray}
where
\begin{equation}
\psi_{1}(t)=\frac{e^{-\frac{3itB}{2}}}{2\sqrt{2}},\nonumber
\end{equation}
\begin{equation}
    \psi_{2}(t)=\frac{e^{-i\frac{1}{2}t(4\sqrt{2} J+B)}[2+\sqrt{2}-(\sqrt{2}-2)e^{4i\sqrt{2}Jt}]}{8\sqrt{2}},\nonumber
\end{equation}
\begin{equation}
  \psi_{3}(t)=\frac{e^{-i\frac{1}{2}t(4\sqrt{2} J+B)}[1+\sqrt{2}-(\sqrt{2}-1)e^{4i\sqrt{2}Jt}]}{4\sqrt{2}},\nonumber
\end{equation}
\begin{equation}
  \psi_{4}(t)=\frac{1}{4}e^\frac{itB}{2}(\sqrt{2}\cos{\frac{4Jt}{\sqrt{2}}}-i\sin{\frac{4Jt}{\sqrt{2}}}),\nonumber
\end{equation}
\begin{equation}
  \psi_{5}(t)=\frac{e^{-i\frac{1}{2}t(4\sqrt{2} J+B)}[2+\sqrt{2}-(\sqrt{2}-2)e^{4i\sqrt{2}Jt}]}{8\sqrt{2}},\nonumber
\end{equation}
\begin{equation}
  \psi_{6}(t)=\frac{1}{4}e^\frac{itB}{2}(\sqrt{2}\cos{\frac{4Jt}{\sqrt{2}}}-2i\sin{\frac{4Jt}{\sqrt{2}}}),\nonumber
\end{equation}
\begin{equation}
  \psi_{7}(t)=\frac{1}{4}e^\frac{itB}{2}(\sqrt{2}\cos{\frac{4Jt}{\sqrt{2}}}-i\sin{\frac{4Jt}{\sqrt{2}}}),\nonumber
\end{equation}
\begin{equation}
  \psi_{8}(t)=\frac{e^\frac{3itB}{2}}{2\sqrt{2}},
\end{equation}
with assumption of $J_{ij}=J$ is a positive real number. By solving the equations, we find that when
\begin{equation}\label{Eq_rela}
B=\frac{4J(1-4C_{1})}{\sqrt{2}(2C_{1}-2C_{2}-1)},~~C_{2}\geq C_{1}\in Z,
\end{equation}
the triatomic complete graph state with a global phase $-i$ can be realized at $t=\sqrt{2}\pi[1+2(C_{2}-C_{1})]/(4J)$, as shown Fig.~\ref{Constent}(a). However, this is just a particular set of solutions. When considering other global phases, there may exist more solutions. As shown in Fig.~\ref{Constent}(b), it is obtained by scanning the population of the target state under different parameters. We can find that three of the peaks correspond exactly to our analytical results, while more peaks also exist. However, in the case of a system with more than three qubits, such a solution is non-existent, and the application of the time-dependent magnetic field with quantum optimal control is necessary.

\section{The multi-selectivity of evolution time}\label{app1}
Within this protocol, populations in the target state exhibit multiple peaks corresponding to different values of $T$. Taking $N=3$ as an example, Fig.~\ref{Multi_T} presents the evolution results with $T=\{0.141, 0.42, 0.696\}~\mu{\rm s}$, respectively. We can find that under these three evolution times, the population of the target state will reach the peak which corresponds to three good optimization results. As we know, a shorter evolution time is more beneficial for the suppression of atomic decoherence. Thus, after comprehensively considering the evolution time, the population of the target state, and the strength of the magnetic field, we choose the parameters of the first peak for simulation in the main text.

\bibliography{Cluster state back-1}
\end{document}